\documentclass[1p]{elsarticle}
\usepackage{physics,bm}
\usepackage{graphicx}
\usepackage{gnuplot-lua-tikz}

\newcommand{\BM}{\begin{pmatrix}}
\newcommand{\EM}{\end{pmatrix}}

\bmdefine{\bx}{x}
\bmdefine{\by}{y}
\bmdefine{\bz}{z}
\bmdefine{\bl}{\lambda}
\bmdefine{\bn}{n}
\renewcommand{\d}{\dagger}
\newcommand{\Lc}{\mathcal{L}}
\newcommand{\Mc}{\mathcal{M}}
\newcommand{\Hh}{\Hat{H}}
\newcommand{\Qh}{\Hat{Q}}
\newcommand{\Ph}{\Hat{P}}
\newcommand{\intx}{\int\!\!d^3x \;}
\newcommand{\psih}{\Hat{\psi}}
\newcommand{\phih}{\Hat{\varphi}}
\newcommand{\ex}{\mathrm{ex}}
\newcommand{\z}{\mathrm{z}}
\newcommand{\f}{\mathrm{f}}
\newcommand{\brasub}{{\!\!\phantom{\bra{}}}}

\begin{document}

\begin{frontmatter}

\title{Thermal and quantum fluctuations of confined Bose-Einstein
 condensate beyond the Bogoliubov approximation}

\author[densi,nagano]{Y.~Nakamura}
\ead{yusuke.n@asagi.waseda.jp}
\author[densi]{T.~Kawaguchi}
\ead{pionelish30@toki.waseda.jp}
\author[densi]{Y.~Torii}
\ead{torii0139@asagi.waseda.jp}
\author[densi]{Y.~Yamanaka}
\ead{yamanaka@waseda.jp}

\address[densi]{Department of Electronic and Physical Systems, Waseda
University, Tokyo 169-8555, Japan} 
\address[nagano]{Nagano Prefectural Kiso Seiho High School, Nagano 397-8571, Japan}

\begin{abstract}
The formulation for zero mode of a Bose-Einstein condensate beyond the
 Bogoliubov approximation at zero temperature 
[Y.Nakamura {\it et al.}, Phys. Rev. A {\bf 89} (2014) 013613] is extended to 
finite temperature. Both thermal and quantum fluctuations are considered in a manner 
consistent with a concept of spontaneous symmetry breakdown for a finite-size system. 
Therefore, we need a proper treatment of the zero mode operators, which 
invoke non-trivial enhancements in  depletion condensate and thermodynamical
quantities such as the specific heat. The enhancements are visible in the
weak interaction case. Our approach reproduces the results of a homogeneous system in the
Bogoliubov approximation in a large particle number limit.
\end{abstract}


\end{frontmatter}

\section{Introduction}
The Bose--Einstein condensate (BEC) system of confined cold atomic 
gas \cite{Cornell,Ketterle,Bradle} has been the most suitable target to 
inspect theories of quantum many-body systems. Quantum field theory, 
which is the most fundamental quantum many-body problem, successfully 
describes the condensation based on a concept of a spontaneous symmetry breakdown 
(SSB) in homogeneous system. Because of spontaneous breakdown
of continuous symmetry, there is always a special excitation mode with zero energy, 
called Nambu--Goldstone mode or zero mode \cite{NGtheorem1,NGtheorem2}. While
most attention is focused on the gapless property of the excitation 
spectrum branch to which the zero mode belongs, we should note that the 
zero mode itself plays a physical role in creating and retaining the ordered 
state associated with SSB. However, mainly because it causes an infrared 
divergence, it is often neglected under the Bogoliubov approximation for the 
homogeneous system, where the zero mode is just a point on the continuous 
spectrum branch and is believed not to affect physical situation very much.

On the other hand, the definite treatment of the zero mode is required for 
inhomogeneous systems, such as trapped systems of cold atomic gases, because the excitation energy is discrete. It is customary 
to specify a particle picture by the unperturbed Hamiltonian that contains 
the first and second powers of the field operator. Then the excitation modes, 
both  zero and Bogoliubov modes, are described by the Bogoliubov--de Gennes (BdG) 
equations \cite{Bogoliubov, deGennes, Fetter}. One is not allowed to neglect 
the zero mode because it appears as a term in a summation, but not as a point 
in an integral. In addition, to neglect the zero mode would break the canonical commutation 
relations of the field, thus implying that the original Heisenberg equation 
and transformation property would be modified.

For the ordinary choice of the unperturbed Hamiltonian up to the second power 
of the field, the ordinary zero mode sector is not diagonalizable in terms of the 
particle operators, but is represented by a pair of quantum coordinates 
$\Qh$ and $\Ph$ \cite{Lewenstein, Matsumoto2002, Mine}. It takes the form of 
free particle as $I\Ph^2/2\,$ with an effective mass $I^{-1}$, and its spectrum 
is continuous. Then, the ground eigenstate gives infinite $\ev*{\Qh^2}$, 
which is a type of infrared divergence.

Recently, we have proposed a new formulation for the trapped system of BEC 
to treat the zero mode operators properly \cite{NTY}, called the interacting
zero mode formulation (IZMF), which is free from the divergence. It is done by 
including the higher power components of the zero mode 
operators, namely the zero mode interactions, into the unperturbed Hamiltonian.
Since then the energy spectrum is discrete, we have obtained a unique 
vacuum that gives both finite $\ev*{\Qh^2}$ and $\ev*{\Ph^2}$. The formulation 
in the previous work was restricted to zero temperature and 
is now extended to finite temperature. We also simultaneously consider 
some fluctuations of the quantum field as the expectations of its products in a 
self-consistent manner and consistently with the Nambu--Goldstone theorem. Our 
condensate system is described by a simultaneous set of three equations including the Gross--Pitaevskii (GP), BdG, and zero mode equations. IZMF also found its application in nuclear
physics, explaining $\alpha$ cluster states 
(including the Hoyle state) in $^{12}$C \cite{NTYO}.

In this paper, following the formulation explained above, we study the averages 
of physical quantities over the unperturbed states such as the depletion
of condensate and specific heat. Thus our considerations 
in this paper is confined to give only 
the unperturbed representation of quantum field theory, 
based on which systematic calculations of higher order corrections from interaction
 should be developed. 
The calculations at the unperturbed level
show that the effects from 
the zero mode operators are found clearly. 
Note though that the choice of the 
unperturbative representation reflects the interaction through renormalization 
and that the zero mode interaction is already included in our unperturbed
representation.

We will find non-trivial enhancements in the calculated quantities for a weaker interaction,
compared with the Bogoliubov approximation at the unperturbed level,
which corresponds to the vestige of the infrared divergence.
Most recent experiments 
to create the condensate phase have been targeting rather strongly 
interacting system, and there are numerical studies of nonperturbative approaches
such as  Monte Carlo simulations \cite{Krauth, Holzmann, DuBois}. 
The results in our present study  cannot be compared directly 
with the experimental datas and those of the simulations. 
Our analysis suggests that  a weakly interacting system
 at low temperatures far from the transition temperature, for which our results are valid,
would be also intriguing.

This paper is organized as follows. In Sect.~\ref{sect:Formulation}, 
we extend our previous formulation (IZMF) \cite{NTY} 
to the case of finite temperature.
In Sect.~\ref{sect:Extension}, we clarify the physical implications of the 
zero mode operators and evaluate 
the depletion of condensate numerically. The energy spectra are estimated analytically
in variational and WKB methods to help us interpret the numerical results.
Next, a general calculational scheme of thermodynamical quantities 
from the partition function is presented in Sect.~\ref{sect:ThermodyanmicalPartition}.
As an example, the specific heat is calculated numerically, and the commitment of the zero
mode to it is examined. The approach of IZMF to the formulation in the Bogoliubov approximation
in a limit of a large particle number is also argued.
Sect.~\ref{sect:Summary} is devoted to the summary.

\section{Formulation of interacting zero mode formulation}\label{sect:Formulation}
We start with the following Hamiltonian to describe the confined Bose atomic gas:
\begin{equation}\label{eq:Hamiltonian}
	\Hh = \intx \left[ \psih^\d \left( h_0 -\mu \right) \psih
	+ \frac{g}{2} \psih^\d\psih^\d\psih\psih\right] \,,
\end{equation}
here $h_0 = -\nabla^2/2m + m\omega^2{\bm x}^2/2$, where $\mu$, $m$, $g$, 
and $\omega$ represent the chemical potential, the mass of an atom, the 
coupling constant, and the strength of the harmonic confinement potential, respectively. 
The Hamiltonian is invariant under the global phase transformation, 
$\psi\, \rightarrow \, e^{i \gamma }\psi$\,. We set $\hbar=1$ throughout this paper.
The bosonic field operator $\psih$ obeys the equal-time canonical commutation relation
\begin{equation}
	\comm{\psih(\bx,t)}{\psih^\d(\bx', t)} = \delta(\bx-\bx') \,.
\end{equation}
On the premise of the spontaneous breakdown of the global phase symmetry, 
the field operator $\psih$ is divided into an order parameter $\xi$ and 
an operator $\phih$ as $\psih = \xi + \phih$ based on the criterion that $\ev*{\psih}=\xi$. 
Note that the definition of the thermal average $\ev{\bullet}$ is implicit at this
stage and is given later in an explicit form. The Hamiltonian (\ref{eq:Hamiltonian}) 
is rewritten in terms of $\phih$ as
\begin{equation}
	\Hh = \Hh_1 + \Hh_2 +\Hh_{3,4} \,,
\end{equation}
where
\begin{align}
	\Hh_1 &= \intx \left[(\phih+\phih^\d)(h_0 -\mu+g\xi^2)\xi\right]\,,\\
	\Hh_2 &= \frac12\intx \BM \phih^\d \,-\phih\EM
		T_0
		\BM \phih \\\phih^\d\EM \,,\\
	\Hh_{3,4} &= \frac g2\intx \left[2\xi(\phih^\d\phih^\d\phih 
				+ \phih^\d\phih\phih)
				+ \phih^\d\phih^\d\phih\phih\right]\,,
\end{align}
with $\Lc_0 = h_0 - \mu + 2g\xi^2$,\,$ \Mc_0 = g\xi^2 $,\, and
\begin{align}
	T_0 = \BM \Lc_0 & \Mc_0 \\ -\Mc_0 & -\Lc_0 \EM \,.
\end{align}
Here and hereafter, $\xi$ is assumed to be a real, an isotropic, and 
a time-independent function.

We take the following GP equation for the order parameter $\xi$, adopted in the 
Hartree--Fock--Bogoliubov approximation \cite{Griffin-HFB-GP}:
\begin{equation} \label{eq:GP}
	\left[ h_0 - \mu + g\left(\xi^2 + 2\ev*{\phih^\d\phih} 
		+ \ev*{\phih\phih}\right) \right]\xi = 0 \,,
\end{equation}
and its squared norm gives the condensate population $N_0=\intx \xi^2$.
The expectations $\ev*{\phih^\d\phih}$ and $\ev*{\phih\phih}$ are parameters in
the counter terms from a standpoint of quantum field theory, as will be explicit later, 
and bear some effects of finite temperature and quantum fluctuations of the 
zero and Bogoliubov modes.
In conjunction with the Hamiltonian ${\hat H}$\,, 
we employ the following matrix in the BdG equation,
\begin{equation} \label{eq:T}
	T = \BM \Lc & \Mc \\ -\Mc^\ast & -\Lc \EM \,,
\end{equation}
with
\begin{equation} \label{eq:LcMc-sHFP}
	\Lc = \Lc_0 +  2g\ev*{\phih^\d\phih} \,,\qquad
	\Mc = \Mc_0 -  g\ev*{\phih\phih}\,.
\end{equation}
Note that the minus sign of $ g\ev*{\phih\phih}$ in $\Mc$
is contrary to the plus sign in the conventional
Hartee--Fock--Bogoliubov approximation. Equation (\ref{eq:LcMc-sHFP}) is the 
BdG equation in the self-consistent Hartree--Fock--Popov 
approximation \cite{Griffin-HFB-GP,GriffinBook} or in the conserving gapless 
one \cite{Kita1,Kita2}, which is consistent with the Nambu--Goldstone 
theorem \cite{NGtheorem1,NGtheorem2}, keeping the zero mode gapless at finite 
temperature as well as at zero temperature. It is straightforward to confirm 
that the matrix $T$ yields the eigenfunction with zero eigenvalue,
\begin{equation}
	T \by_0 = \bm{0} \,, \qquad \by_0 \equiv \BM \xi & -\xi\EM^t \,,
\end{equation}
while there is no gapless eigenfunction for $T_0$ in the conventional 
Hartee--Fock--Bogoliubov approximation. However, it is favorable 
that $T$ has the same matrix symmetry as $T_0$:
\begin{equation}
	\sigma_1 T \sigma_1 = -T  \,,\qquad
	\sigma_3 T \sigma_3 = T^\d  \,,
\end{equation}
where $\sigma_i$ denotes the $i$-th Pauli's matrix.
Thus, we obtain an orthonormal set to expand the field operator similar to 
the case of zero temperature \cite{Lewenstein,Matsumoto2002} as
\begin{align}\label{eq:completeset}
	\sum_\bl  \left[ \by_\bl(\bx) \by_\bl^\d(\bx') 
		- \bz_\bl(\bx) \bz_\bl^\d(\bx') \right] 
	+\by_0(\bx)\by_{-1}^\d(\bx') + \by_{-1}(\bx)\by_{0}^\d(\bx')  
		= \sigma_3 \delta(\bx-\bx') \,.
\end{align}
Here, $\by_\bl$ and $\bz_\bl$ are the eigenfunctions of $T$ which belong to 
the eigenvalues $\omega_\bl$ and $-\omega_\bl$, respectively, with a set of quantum 
numbers $\bl=(n,\ell,m)\,$
and are related to each other by $\by_\bl = \sigma_1 \bz_\bl^*$,
\begin{equation}\label{eq:BdG}
T \by_\bl= \omega_\bl \by_\bl\,, \quad T \bz_\bl=- \omega_\bl \bz_\bl \,.
\end{equation}
The function $\by_{-1}$ denotes the adjoint mode that is given by
\begin{equation}
	\label{eq:def_y-1}
	\by_{-1} = \BM \eta & \eta \EM^t \,,\qquad (\Lc+\Mc)\eta = I \xi \,.
\end{equation}
The factor $I$ is a normalization constant adjusted to satisfy 
$\intx \eta(\bx) \xi(\bx) = 1/2$\,. When $\ev*{\phih^\d\phih}$ and 
$\ev*{\phih\phih}$ are small and negligible, we have $\eta=d\xi/dN_0$ and 
$I = d\mu/dN_0$ by differentiating the GP equation with respect to 
$N_0$ \cite{Kobayashi2009}.
The field operator is now expanded as $\phih = \phih_{z} + \phih_{\ex}$ where
\begin{align} \label{eq:fieldexpansion}
	\phih_\z = -i \Qh \xi + \Ph \eta \,,\quad
	\phih_{\ex} = \sum_{\bl} \left[ \hat a_\bl u_\bl 
		+ \hat a_\bl^\d v_\bl^*\right] \,,
\end{align}
with $\by_\bl = \bigl( u_\bl \quad v_\bl \bigr)^t$, which is normalized as
$\intx (|u_\bl|^2 - |v_\bl|^2)=1$\,.
Note that while $\phih$ satisfies the canonical commutation 
relation $\comm{\phih(\bx)}{\phih^\d(\bx')}=\delta(\bx-\bx')$, 
but $\phih_\ex$ does not, which is a definite
reason why one may not drop the zero mode operators for inhomogeneous
system. The commutation relations of $\phih$ gives
\begin{equation} \label{eq:CCRofQP}
	\comm{\Qh}{\Ph} = i\,,\qquad
	\comm{\hat a_\bl}{\hat a_{\bl'}^\d} = \delta_{\bl\bl'} \,,
\end{equation}
and otherwise the vanishing ones.
Here, $\Qh$ and $\Ph$ are Hermitian and called the zero mode operators.

Considering $\ev*{\phih^\d\phih}$ and $\ev*{\phih\phih}$, 
one can introduce $\Hh_u'$ as a candidate for the unperturbed Hamiltonian:
\begin{align}\label{eq:Huprime}
 	\Hh_u' &= \Hh_1 + \Hh_2 + \delta\Hh \,,\\
 	\delta\Hh &=\intx\left[
	 2g\ev*{\phih^\d\phih} \bigl\{ \xi\left( \phih+\phih^\d \right) 
	+ \phih^\d\phih \bigr\}
	 +g\ev*{\phih\phih} \bigl\{ \xi\left( \phih+\phih^\d \right) 
	-\left(\phih\phih+ \phih^\d\phih^\d\right)/2 \bigr\}\right]\,.
\end{align}
As mentioned, $\delta\Hh$ acts as the counter terms 
[see Eqs.~(\ref{eq:GP}) and (\ref{eq:LcMc-sHFP})], 
and eliminates the first power term of $\phih$ from $\Hh_u'$. We thus have
\begin{equation}\label{eq:Huprime2}
 	\Hh_u' = \frac12\intx \BM \phih^\d -\phih\EM T \BM \phih \\ \phih^\d\EM \,,
\end{equation}
or in terms of $a_\bl$, $\Qh$, and $\Ph$ as
\begin{equation} \label{eq:H2}
	\Hh_u' = \frac{I\Ph^2}{2} + \sum_{\bl} \omega_\bl \hat a_\bl^\d \hat a_\bl \,.
\end{equation}
The zero mode part of this Hamiltonian is not diagonal in terms of the 
annihilation- and creation-operators, but is the free particle one with a 
continuous spectrum. While it seems natural to choose $\Hh_u'$ as the 
unperturbed Hamiltonian on the premise of small $\phih$, it involves a fatal 
problem as is discussed in Ref.~\cite{NTY}. Because of the free-particle form 
of the zero mode part in $\Hh_u'$, the quantity $\ev*{\Qh^2}$ and consequently 
the number density $\ev*{\psih^\d\psih}$ diverge. To eliminate the divergence, 
it had been proposed to replace $\psih = \xi + \phih_z + \phih_\ex$ with a new 
expression of the field as \cite{Lewenstein,Matsumoto2002}
\begin{equation}  \label{eq:LewensteinAnsatz}
	\psih_{\rm LY}  \equiv \bigl( \xi + \eta \Ph + \phih_\ex \bigr)e^{-i\Qh}
\,,
\end{equation}
both being identical with each other only for small $\Qh$. Although 
$\ev*{\psih^\d_{\rm LY}\psih_{\rm LY}}$ is free from the divergence, one 
encounters new problems: First, the quantity $\ev*{\Qh^2}$ still diverges contrary 
to the assumption of small $\Qh$. Second, the operator $\psih_{\rm LY}$ 
does not fulfill the canonical commutation relation, which is the very 
foundation of the quantum filed theory.

The zero mode excitations on the unperturbed steady state accumulate very 
easily due to the zero energy nature. The accumulation brings the infrared 
divergence above and also invalidates the choice of the unperturbed 
Hamiltonian up to the second power of the zero mode operators because 
the expectations of the operators with their products are not small in general. 
We take note of the fact that the interaction term in the total Hamiltonian,  
$g\psih^\d\psih^\d\psih\psih/2$, has zero mode interaction terms, and we
have proposed a new unperturbed Hamiltonian \cite{NTY}, which contains not 
only the first and second powers of the zero mode operators but also the 
higher ones. Thus, the zero mode interaction is introduced at the unperturbed 
level, and we refer to the new formulation as the IZMF. While it has been done at zero temperature 
and without the counter terms 
$\ev*{\phih^\d\phih}$ and $\ev*{\phih\phih}$ in Ref.~\cite{NTY}, we now extend it to 
finite temperature and introduce the unperturbed Hamiltonian as follows:
\begin{align}\label{eq:Hu1}
	\Hh_u &= \Hh_1 + \Hh_2 + \left[ \Hh_{3,4} \right]_{QP} 
		- \delta\mu \Ph -\delta\nu\Qh + \delta\Hh \,.
\end{align}
Here, the symbol $[\cdots]_{QP}$ indicates that all the terms consisting only 
of the zero mode operators are picked up. The counter terms $\delta\mu$ and 
$\delta\nu$ are chosen to hold $\ev*{\Qh}=\ev*{\Ph}=0$, which is necessary 
for the criterion of division $\ev*{\phih}=0$.

In this paper, we consider the thermal average 
$\mathrm{Tr}\,[\hat \rho\, \hat{O}]/\mathrm{Tr}\,[\hat\rho]$ 
over the equilibrium density matrix 
\begin{equation}
\hat\rho = \exp\,( -\beta \Hh_u )\,,
\label{eq:DensityMatrix1}
\end{equation}
defined by the unperturbed Hamiltonian $\Hh_u$, where $\beta$ is the inverse 
temperature. Then, because the thermal average should be time-independent, 
we obtain \linebreak
 $\ev*{[\Qh,\,\Hh_u]}=\ev*{[\Ph,\,\Hh_u]}=0$. 
We can determine the counter terms as
\begin{equation}\label{eq:deltamu}
	\delta\mu = -i\ev{\comm{\Qh}{[\Hh_{3,4}]_{QP}}} \,,\quad
	\delta\nu =  i\ev{\comm{\Ph}{[\Hh_{3,4}]_{QP}}} \,.
\end{equation}
In the case where $\xi$ is real, one can show that there is no odd power term 
of $\Qh$ in $\Hh_{3,4}$ so that $\delta\nu = 0$. However, there are
always odd power terms of $\Ph$, and we have to deal with $\delta \mu$. 
Substituting the field expansion (\ref{eq:fieldexpansion})
into Eq.~({\ref{eq:Hu1}), we obtain $\Hh_u = \Hh_{u,\z} + \Hh_{u,\ex}$, where
\begin{align}
	\Hh_{u,\z} &= -(\delta\mu + 4C)\Ph + \frac{I-4D}{2} \Ph^2 
	+ 2B \Qh\Ph\Qh + 2D\Ph^3 	\notag\\
	&+ \frac{1}{2}A\Qh^4 -2B\Qh^2 + C\Qh\Ph^2\Qh + \frac{1}{2}E\Ph^4 \,,\\
	\Hh_{u,\ex} &= \sum_{\bl} \omega_\bl \hat{a}_\bl^\d \hat{a}_\bl \,,
\end{align}
with
\begin{alignat}{3} \label{eq:defAI}
	A &= g\intx \xi^4 \,,\;\;&
	B &= g\intx \xi^3 \eta\,,\;\;&
	C &= g\intx \xi^2 \eta^2\,,\;\; \notag\\
	D &= g\intx \xi \eta^3 \,,\;\;&
	E &= g\intx \eta^4 \,.\;\;&
\end{alignat}

We set up the eigenequation for $\Hh_{u,\z}$,
\begin{equation}
\Hh_{u,\z}\ket{\Psi_\nu}_\z = E_\nu \ket{\Psi_\nu}_\z \,,
\label{eq:ZeroModeEq}
\end{equation}
which we refer as the zero mode equation. The eigenstate space for $\Hh_{u,\ex}$ 
is the Fock space whose element is denoted by
\begin{equation}
\Hh_{u,\ex}\ket{\{n_\bl\}}_\ex = \sum_{\bl} n_\bl \omega_\bl
 \ket{\{n_\bl\}}_\ex \,.
\end{equation} 
The eigenstate for $\Hh_u$ is expressed as a direct product of the above ones,
\begin{equation}
\Hh_{u}\ket{\nu, \{n_\bl\}}= \left(E_\nu+ \sum_{\bl} n_\bl \omega_\bl\right)
\ket{\nu, \{n_\bl\}} \,,
\end{equation}
where $\ket{\nu, \{n_\bl\}} =\ket{\Psi_\nu}_z \otimes \ket{\{n_\bl\}}_\ex$.
The density matrix in Eq.~(\ref{eq:DensityMatrix1}) is 
\begin{align}
\hat\rho = \left( \sum_\nu e^{-\beta E_\nu}\ket{\Psi_\nu}_\z {\brasub}_\z\bra{\Psi_\nu}
\right)
\otimes \prod_{\bl} \left( \sum_{n_\bl} e^{-\beta n_\bl \omega_\bl  }
\ket{n_\bl}_\ex {\brasub}_\ex\bra{n_\bl}   \right)\,.
\label{eq:DensityMatrix2}
\end{align}
Note that because $\delta\mu$, given by the thermal average (\ref{eq:deltamu})}, 
is also included in $\Hh_{u,\z}$ it should be determined self-consistently.

\section{Zero mode operators and depletion of condensate}\label{sect:Extension}
In this section, we calculate some quantities concerned with the particle numbers, 
following the IZMF presented in the previous section. 
The calculational steps are as follows: With the fixed total particle number 
$N = \intx \ev*{\psih^\d\psih}$ and considering that 
$\ev*{\phih^\d\phih}$ and $\ev*{\phih\phih}$ are given, one can obtain 
$\xi$, $\mu$, $\eta$, and $I$ from Eqs.~(\ref{eq:GP}) and (\ref{eq:def_y-1}). 
Next, one obtains $\by_\bl$ and $\omega_\bl$ by solving the BdG equation, 
and $\ket{\Psi_\nu}$ and $E_\nu$ by solving the zero mode 
equation~(\ref{eq:ZeroModeEq}). Then, the thermal averages 
$\ev*{\phih^\d\phih}$ and $\ev*{\phih\phih}$
can be calculated. These steps should be done in a self-consistent manner.

The total number $N$ is expressed in our unperturbed calculations as
\begin{align}
N&=N_0 + N_\f \notag \\
N_0& = \intx \xi^2 \notag \\
N_\f&= \intx \ev{\phih^\dagger \phih} \notag \\
&=
 \intx \left[\ev{\Qh^2} \xi^2+ \ev{\Ph^2}\eta^2 
+ \sum_{\bl}\left\{ \ev{\hat a_\bl^\d \hat a_\bl}
 (|u_\bl|^2+|v_\bl|^2)+|v_\bl|^2\right\}\right]\,.
\end{align}
where $\ev*{\Qh}=\ev*{\Ph}=0$ has been used. The averages are explicitly
\begin{align}
\ev{\Qh^2} &=  \sum_\nu e^{-\beta E_\nu} {\brasub}_\z\bra{\Psi_\nu}
{\Qh^2}\ket{\Psi_\nu}_\z/ 
\sum_\nu e^{-\beta E_\nu} \notag \\
\ev{\Ph^2}  &=  \sum_\nu e^{-\beta E_\nu}{\brasub}_\z\bra{\Psi_\nu}
{\Ph^2}\ket{\Psi_\nu}_\z/ 
\sum_\nu e^{-\beta E_\nu} \notag \\
\ev{\hat a_\bl^\d \hat a_\bl} & =  \frac{1}{e^{\beta \omega_\bl}-1} \,.
\end{align}
The presence of the zero mode contribution, $\ev*{\Qh^2}$ and $\ev*{\Ph^2}$, 
is characteristic of our approach.

\subsection{Interpretation of zero mode operators}
In order to elucidate the physical meanings of zero mode operators, 
we rewrite $\psih$ as 
\begin{align} \label{eq:xi_functional}
\psih \simeq \left( 1  -i\Qh+ \Ph \frac{d}{dN_0} \right)\xi[\theta_0,N_0] 
+ \phih_\ex  \simeq \xi\bigl[ \theta_0- \Qh,\, N_0 + \Ph \bigr] + \phih_\ex  \,,
\end{align}
on a premise of small $\ev*{\Qh^2}$ and $\ev*{\Ph^2}/N_0^2$. We denote 
explicitly the squared norm $N_0$ and the global phase $\theta_0$ of 
$\xi$, where $\theta_0$ has been set to zero in this paper.
Then, it is natural to interpret $\Qh$ and $\Ph$ as the displacement operators of
the global phase and the number of condensate atoms, respectively. 
Note that this interpretation is valid only if the variances of $\Qh$ and $\Ph$ 
are sufficiently small. While the global phase is $\theta_0$, it fluctuates 
with $\Delta Q = \ev*{\Qh^2}^{1/2}$. Likewise, the number of condensate atoms 
fluctuates with $\Delta P = \ev*{\Ph^2}^{1/2}$. The canonical commutation 
relation (\ref{eq:CCRofQP}) immediately leads the uncertainty relation
$\Delta Q \Delta P \ge 1/2$ at zero temperature, and the lowest limit 
there is pushed up at finite temperature \cite{MRUY}.
Although Eq.~(\ref{eq:xi_functional}) looks similar to  
Eq.~(\ref{eq:LewensteinAnsatz}) at first glance, we have to pay 
attention to both of them being different: while Eq.~(\ref{eq:xi_functional}) is 
just an approximate expression to interpret the physical meanings of the 
zero mode operators, Eq.~(\ref{eq:LewensteinAnsatz}) 
is an exact definition in terms of new operators.

\subsection{Variational estimation}
Before performing numerical calculations, 
let us estimate the zero mode effects in the case of a large $N_0$ limit by 
evaluating $E_0$ and $E_1$ variationally. We set the trial functions as
\begin{align}
	\bra{Q}\ket{\Psi_0}_\z &= \left( \frac{1}{2\pi \alpha_0^2}\right)^{1/4}\, 
	e^{-{Q^2}/{4\alpha_0^2}} \,,\\
	\bra{Q}\ket{\Psi_1}_\z &= \left( \frac{1}{2\pi \alpha_1^6}\right)^{1/4}\,q\, 
	e^{-{Q^2}/{4\alpha_1^2}} \,,
\end{align}
in $Q$-representation ($\Qh \ket{Q} = Q \ket{Q}$) with variational 
parameters $\alpha_\nu$. Only the terms proportional to $A$ and $I$ are dominant
in the large $N_0$ limit, as is discussed in Ref.~\cite{NTY}, and we obtain
\begin{align}
	\mathcal{E}_0
	\simeq \frac32 A \alpha_0^4 + \frac{I}{8\alpha_0^2}\,,\qquad
	\mathcal{E}_1
	\simeq \frac{15}2 A \alpha_1^4 + \frac{3I}{8\alpha_1^2}\,,
\end{align}
where $	\mathcal{E}_\nu = {\brasub}_\z \bra{\Psi_\nu}{\Hh_{u,\z}} 
\ket{\Psi_\nu}_\z$. The variational parameters that minimize 
$\mathcal{E}_\nu$ are $\alpha_0 = \sqrt[6]{I/24A}$ and
$\alpha_1 = \sqrt[6]{I/40A}$, and the first excitation energy
$\Delta \mathcal{E} \equiv \mathcal{E}_1 -\mathcal{E}_0$ is approximately 
$1.38\sqrt[3]{AI^2}$. To evaluate the values of $A$ and $I$ roughly, we put 
$\ev*{\phih^\d\phih}=\ev*{\phih\phih}=0$
and apply the Thomas--Fermi approximation to the GP equation \cite{PethickSmith}.
Using Eq.~(\ref{eq:defAI}) and the relation $I = d\mu/dN_0$, we finally estimate
\begin{equation}
	\Delta\mathcal{E} = 0.92 \,  \omega \,
	\left(\frac{a_{\mathrm{s}}}{a_{\mathrm{osc}}}\right)^{2/5} \,N_0^{1/15} \,,
	\label{eq:DelE}
\end{equation}
where $a_{\mathrm{s}}$ is the $s$-wave scattering length related to the coupling constant
by $g=4\pi a_{\mathrm{s}}/m$, and $a_{\mathrm{osc}}$ is the characteristic length
for the harmonic oscillator given by $1/\sqrt{m\omega}$.

First $\Delta\mathcal{E}$ diverges in the limit of infinite $N_0$, and then the zero mode 
excitation is strongly suppressed. However, it cannot be neglected in 
realistic experiments of cold atoms because the exponent in the power of $N_0$ 
in Eq.~(\ref{eq:DelE}) is very small. A typical ratio of $a_{\mathrm{s}}/a_{\mathrm{osc}}$ 
is the order of $10^{-2}$, so even for condensates of a very small number of 
atoms, say $N_0=1$, and with a very large one, say $N_0=10^{10}$, 
$\Delta\mathcal{E}$ is still the order of $\omega$. Note also that, 
$\Delta\mathcal{E}$ reduces to zero in the weak interaction limit 
$a_\mathrm{s}\to 0$, and the discrete spectrum $\mathcal{E}_\nu$ approaches 
a continuous one and the wavefunction $\braket{Q}{\Psi_\nu}_\z$ spreads wide 
in $Q$-space. In other words, the non-linear unperturbed Hamiltonian (\ref{eq:Hu1}) 
returns to the bilinear one (\ref{eq:H2}). 

In the case of sufficiently low temperature 
$\beta\Delta\mathcal{E} \gg 1$, $\Delta Q$ and $\Delta P$
are estimated as $\alpha_0$ and $1/2\alpha_0$, respectively. 
It implies $\Delta Q \sim N_0^{-1/3}$ and $\Delta P/N_0 \sim N_0^{-2/3}$, 
both of which vanish in the limit of infinite $N_0$.
This consequence is different from the one based on the use of a coherent state, 
$\ket{\theta_0, N_0}$, for which we have $\hat{a}_0 \ket{\theta_0, N_0} = 
\sqrt{N_0}\,e^{i\theta_0}\,\ket{\theta_0, N_0}\,,$
with $[\hat{a}_0,\,\hat{a}_0^\d]=1$. As is well-known, it leads a number 
fluctuation of $N_0^{1/2}$, while our number fluctuation $\Delta P$ has a smaller 
exponent, $1/3$. Basically, the zero mode interaction reduces 
the number fluctuation, whereas it enhances the phase fluctuation more than 
that for the naive coherent state. We emphasize that the operators 
$\hat{a}_0$ and $\hat{a}_0^\dagger$ cannot diagonalize the zero mode part of 
the bilinear Hamiltonian [see Eq.~(\ref{eq:Hu1})].

\subsection{WKB estimation}
The variational method in the previous subsection gives only the two lowest 
eigenvalues, $\mathcal{E}_0$ and $\mathcal{E}_1$. When temperature increases, 
contributions from higher energy states have to be considered. Higher 
eigenvalues can be estimated in the WKB approximation. As in the previous 
subsection, it is assumed that  $\Hh_{u,\z}$ is $I \Ph^2/2 + A\Qh^4/2$ 
approximately. Then, we derive in the WKB approximation,
\begin{align}
&\int_{-q_\nu}^{q_\nu} \sqrt{\frac{2}{I}\left(E_\nu-\frac 12 A Q^4\right)}\, dQ
=\pi \left(\nu+\frac 12 \right) \notag \\
& \qquad \left( \nu=0,1,2,\cdots \right) \,,
\end{align}
where $\pm q_\nu$ are the turning points, satisfying $E_\nu- A q_\nu^4/2=0$\,.
After the integration, $E_\nu$ is solved as 
\begin{align}
E_\nu \simeq\left\{\frac{\pi}{1.74}\left( \nu+\frac 12\right)\right\}^{4/3}
(AI^2)^{1/3}\,,
\end{align}
and, when the Thomas--Fermi approximation is used to evaluate $A$ and $I$ 
as before, it is approximated as
\begin{align}
E_\nu \simeq 0.58 (2\nu+1)^{4/3}
\left( \frac{a_{\mathrm{s}}}{a_{\mathrm{osc}}}\right)^{2/5}N_0^{1/15}
	\omega \,.
\label{eq:DeltaEnu}
\end{align}
It is common for $E_\nu$ to be proportional to $(AI^2)^{1/3}$, or to 
$\left( {a_{\mathrm{s}}}/{a_{\mathrm{osc}}}\right)^{2/5}N_0^{1/15} \omega$ 
in both of the variational and WKB approximations. Let us define 
$\Delta E_\nu =E_{\nu+1}-E_\nu$. In the large $\nu$ limit, 
$\Delta E_\nu \simeq 2.9 \nu^{1/3} \left( {a_{\mathrm{s}}}/{a_{\mathrm{osc}}}
\right)^{2/5}N_0^{1/15} \omega \propto \nu^{1/3}$\,. 
The WKB approximation is not good for a small $\nu$, but the 
ratio $\Delta E_0/\Delta\mathcal{E}$ is calculated to be approximately 2 
from Eqs.~(\ref{eq:DelE}) and (\ref{eq:DeltaEnu}). Although $E_\nu$ in 
Eq.~(\ref{eq:DeltaEnu}) tends to be large compared with that obtained from 
solving Eq.~(\ref{eq:ZeroModeEq}) numerically, its parameter dependence 
traces the numerical one well.

\subsection{Numerical result for depletion of condensate}
We now show the results of the depletion of the condensate $1 - N_0/ N$  
by numerically solving the coupled system of the GP Eq.~(\ref{eq:GP}), 
BdG Eqs.~(\ref{eq:BdG}) and (\ref{eq:def_y-1}), 
and zero mode Eq.~(\ref{eq:ZeroModeEq}).
Here and hereafter, we fix the total particle number $N$ to $10^3$ 
and vary the interaction strength 
$a_{\mathrm{s}}/a_{\mathrm{osc}}$ from $10^{-4}$ to $10^{-1}$, 
which covers the typical experiments 
\cite{PhysRevLett.77.4984,PhysRevA.88.053614}. The temperature is scaled by 
the critical temperature of ideal gas, 
$k_\mathrm{B}T_c = \omega \sqrt[3]{N/\zeta(3)} \simeq 9.4 \omega$.

\begin{figure}[t]
	\centering
	\includegraphics{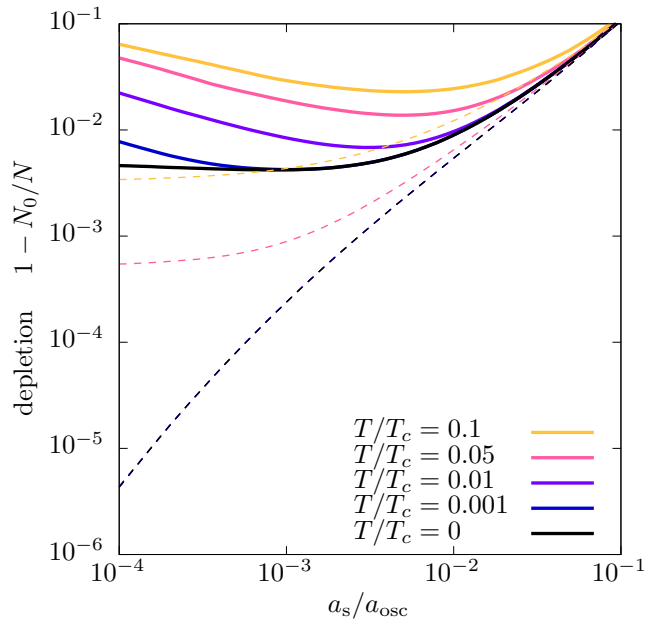}
	\caption{(Color Online) Depletions of the condensate $1 - N_0/ N$ as 
	a function of ${a_{\mathrm{s}}}/{a_{\mathrm{osc}}}$ (solid lines) and
	the same results with the Bogoliubov approximation, where both $\ev*{\Qh^2}$
	 and $\ev*{\Ph^2}$ are set to zero (dashed lines). The dashed lines with 
	 $T/T_c=0.01$ and $0.001$ are closely overlapping to that with $T/T_c = 0$.
	}
	\label{fig:depletions}
\end{figure}

The result of the depletion is illustrated in Fig.~\ref{fig:depletions}. 
The fact is that while the zero mode contribution, absent in the Bogoliubov 
approximation, is negligible for the stronger interaction, it becomes noticeable 
for the weaker interaction. The ${a_{\mathrm{s}}}/{a_{\mathrm{osc}}}$-dependence 
in the weakly interacting case is enhanced as the temperature increases, and then 
the depletion, or the reduction of $N_0$, is attributed mainly to quantum and 
thermal fluctuations of the zero mode.

\begin{figure}[t]
	\centering
  \includegraphics[height=0.6\textheight]{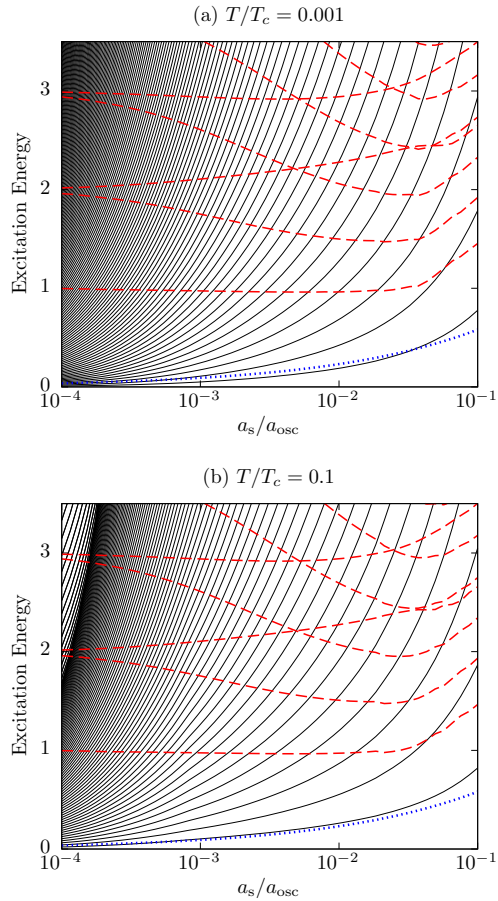}
	\caption{(Color Online) Excitation energy as a function of 
		${a_{\mathrm{s}}}/{a_{\mathrm{osc}}}$ in unit of $\omega$. 
		The dashed red line denotes the BdG excitation $\omega_{\bl}$, and
		the solid black line denotes that of zero 
		mode $E_\nu - E_0$ ($\nu=1, 2, \cdots)$. The dotted blue line 
		indicates the variational result for the first zero mode 
		excitation $\Delta\mathcal{E}$.
	}
	\label{fig:spectra}
\end{figure}

In Fig.~\ref{fig:spectra}, we show the spectra of the zero mode and BdG excitations 
for lower and higher temperatures as a function of the interaction strength, 
$a_{\mathrm{s}}/a_{\mathrm{osc}}$, which give us an account 
of the behaviors in Fig.~\ref{fig:depletions}. The spectra depend only slightly on temperature.
The excitation energy of the zero mode  increases conspicuously
as the interaction becomes stronger, whereas that of the BdG mode is rather robust.
Explicitly, the first excited energy of the zero mode $E_1 - E_0$ is approximately $0.02\omega$ 
for ${a_{\mathrm{s}}}/{a_{\mathrm{osc}}}=10^{-4}$, and increases to $0.8\omega$ for 
${a_{\mathrm{s}}}/{a_{\mathrm{osc}}}=10^{-1}$, which is comparable with 
the BdG excitation energy. The theoretical estimations~(\ref{eq:DelE}) 
and (\ref{eq:DeltaEnu}), predicting that energy levels vary 
as $(a_{\mathrm{s}}/a_{\mathrm{osc}})^{2/5}$, roughly trace the behaviors 
in Fig.~\ref{fig:spectra}. Considering that a typical thermal energy $k_\mathrm{B} T$
varies between $0.001 \omega$ and $\omega$ over the temperature range under
consideration, we see that the BdG mode is hardly excited thermally. 
The thermal excitation of the zero mode is also inhibited for stronger interaction, 
such as for $a_{\mathrm{s}}/a_{\mathrm{osc}}=10^{-1}$.
However, as ${a_{\mathrm{s}}}/{a_{\mathrm{osc}}}$ is smaller,
the spacing of the zero mode excitation energy becomes narrower, and 
much narrower than $k_\mathrm{B} T \sim \omega$ for $T/T_c = 0.1$
and is comparable with $k_\mathrm{B} T \sim 0.01 \omega$ for $T/T_c = 0.001$. 
As a result, the thermal excitations of the zero mode are inconsiderable for $T/T_c = 0.001$, 
but substantial for $T/T_c = 0.1$.

\begin{figure}[t]
	\centering
	\includegraphics{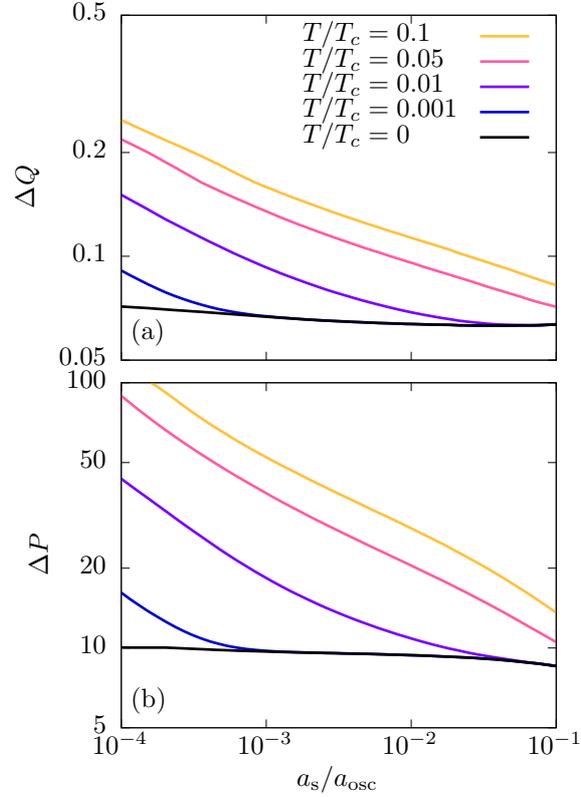}
	\caption{(Color Online) Interaction dependences of the phase 
	fluctuation $\Delta Q$ and number fluctuation $\Delta P$.}
	\label{fig:interactionDep}
\end{figure} %
\begin{figure}[t]
	\centering
	\includegraphics{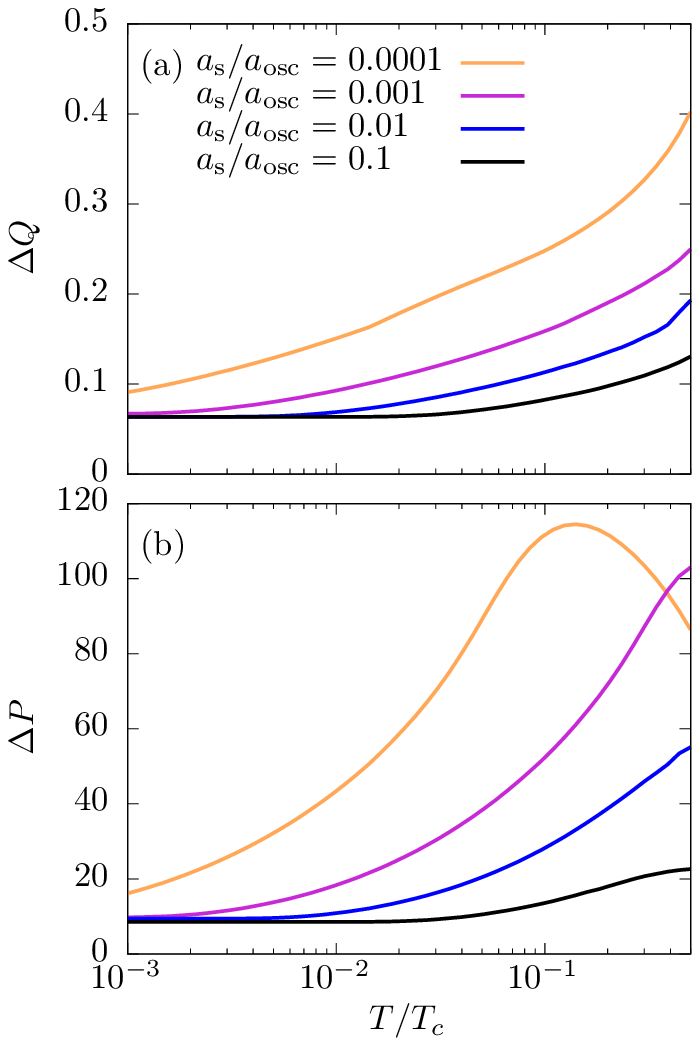}
	\caption{(Color Online) Temperature dependences of the phase 
	fluctuation $\Delta Q$ and number fluctuation $\Delta P$.}
	\label{fig:temperatureDep}	
\end{figure} %

It is important to argue about the fluctuations of the phase $\Delta Q$ and the number 
of condensate atoms $\Delta P$ from the dependences of the zero 
mode excitation spectrum on temperature and interaction strength. Their numerical
results are shown in Figs.~\ref{fig:interactionDep} and \ref{fig:temperatureDep}.
Both of the fluctuations are almost constant irrespective of the interaction strength 
at zero temperature, which is in good agreement with the analytical results in Ref.~\cite{NTY},
and increase with increasing temperature. This is more apparent in a weakly interacting case, 
where the spacing of the zero mode excitation energy is narrow and the thermal
excitation is active. The irregular temperature dependence of $\Delta P$ in the weakly 
interacting case ($a_{\mathrm{s}}/a_{\mathrm{osc}}=10^{-4}$) is seen in 
Fig.~\ref{fig:temperatureDep}, and it does not increase
monotonically, but decreases beyond $T/T_c=0.1$. This is because a smaller 
number of condensate atoms $N_0$ result in a smaller fluctuation $\Delta P$.

\section{Thermodynamical quantity and partition function}\label{sect:ThermodyanmicalPartition}

We consider the influences of the zero mode excitations on thermodynamical quantities
that are derived from the partition function. 
As an example, the specific heat is calculated numerically.

\subsection{Partition function}

The partition function in our approximation is given in a factorized form,
\begin{align} 
Z&=\mathrm{Tr}\,[\exp\,( -\beta \Hh_u )]=Z_\z Z_\ex \notag \\
Z_\z&=  \mathrm{Tr}_{\z}\,[\exp\,( -\beta \Hh_{u,\z})]
= \sum_\nu\, e^{-\beta E_\nu}  \notag \\
Z_\ex&=  \mathrm{Tr}_{\ex}\,[\exp\,( -\beta \Hh_{u,\ex})]
= \prod_\bl\, \frac{e^{\beta \omega_\bl}}
{e^{\beta \omega_\bl}-1} \,,
\end{align}
where $\mathrm{Tr}_{\z}$ and $\mathrm{Tr}_{\ex}$ stand for traces over the zero 
mode subspace and Fock space of the Bogoliubov modes, respectively. Because 
$\ln Z=\ln Z_\z+ \ln Z_\ex$, the thermodynamical quantities are given as sums of 
zero mode contributions and Bogoliubov ones.
For example, the internal energy is
\begin{align}
U & = - \frac{\partial}{\partial \beta} \ln Z=-  \frac{\partial}{\partial \beta}
 \ln Z_\z - \frac{\partial}{\partial \beta} \ln Z_\ex \notag \\
&= \frac{\sum_\nu\, E_\nu\, e^{-\beta E_\nu}}{\sum_\nu\, e^{-\beta E_\nu}} 
+ \sum_\bl\, \frac{\omega_\bl}
{e^{\beta \omega_\bl}-1}\,.
\end{align}

\subsection{Specific heat}
The specific heat is derived from the above $Z$,
\begin{align}
C&= \frac{1}{k_\mathrm{B} T^2} \frac{\partial^2}{\partial \beta^2 }\ln Z
=C_\z+ C_\ex \notag \\
C_\z & =\frac{1}{k_\mathrm{B} T^2}
\left\{ \frac{\sum_\nu\, E^2_\nu\, e^{-\beta E_\nu}}{\sum_\nu\, e^{-\beta E_\nu}} 
-\left(\frac{\sum_\nu\, E_\nu\, e^{-\beta E_\nu}}{\sum_\nu\, 
e^{-\beta E_\nu}}\right)^2  \right\} \notag \\
C_\ex & =\frac{1}{k_\mathrm{B} T^2}
\sum_\bl\, \frac{\omega_\bl^2\, e^{\beta \omega_\bl}}
{(e^{\beta \omega_\bl}-1)^2}\,.
\label{eq:SpecificHeat}
\end{align}

\begin{figure}[t]
	\centering
	\includegraphics{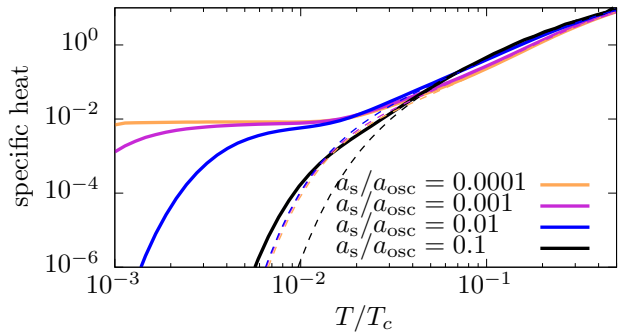}
	\caption{(Color Online) Temperature dependences of the specific heat 
		beyond the Bogoliubov approximation (solid lines) and 
		the one in the Bogoliubov approximation (dashed lines).}
	\label{fig:specificheat}	
\end{figure} %

The numerical result of the specific heat $C$  from 
Eq.~(\ref{eq:SpecificHeat}) is shown in Fig.~\ref{fig:specificheat}. For comparison, we  
recall that the specific heat for an ideal gas trapped by a harmonic potential behaves as 
$(T/T_c)^3$ under the approximation that a sum over the 
spectrum may be replaced with an integral \cite{PethickSmith}. 
The result of the specific heat in the Bogoliubov approximation is 
also depicted in Fig.~\ref{fig:specificheat}. 
Both the plots, in and beyond the Bogoliubov approximation, are roughly fitted by a 
line of the ideal gas approximation $(T/T_c)^3$ above 
$T/T_c \sim 5 \times  10^{-2}$.
Below the temperature, 
the plot in the Bogoliubov approximation (dashed line in Fig.~\ref{fig:specificheat}) 
decreases rapidly due to suppression of the thermal excitation
of the BdG mode, and the repulsive interaction increases the energy level of the BdG mode 
as in Fig.~\ref{fig:spectra} and reinforces the suppression.
A notable result in our approach is that the plot 
beyond the Bogoliubov approximation (solid line in Fig.~\ref{fig:specificheat})
gets a plateau for weaker interaction strength and at lower temperature instead of
falling down immediately. This is true as long as 
the spacing of the zero mode excitation energy (see Fig.~\ref{fig:spectra})
is smaller than a typical thermal energy $k_\mathrm{B} T$.
In other words, an external energy is distributed to the densely populated excitation 
energy levels of the zero mode, thus the specific heat is kept constant for a while.
When the temperature increases, the effects of the BdG modes with many degrees of freedom
dominates those of the zero mode that has a single degree of freedom.

\begin{figure}[t]
	\centering
	\includegraphics{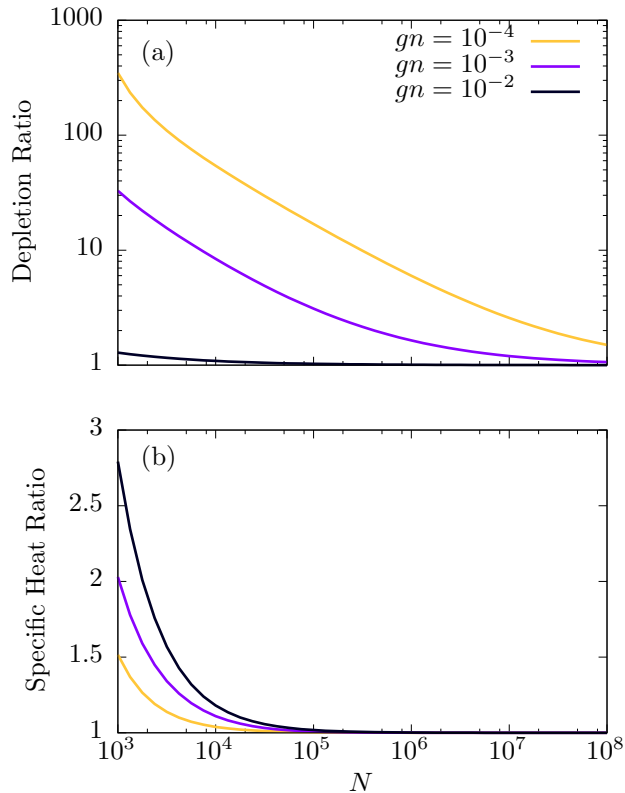}
	\caption{(Color Online) Ratios of the depletion and specific heat in the homogeneous IZMF 
	to those in the Bogoliubov approximation as a function of $N$ at $T/T_c=0.01$. 
	The quantity $gn$ is set dimensionless here.}
	\label{fig:thermoDynamicial}	
\end{figure} %

\subsection{Comparison between IZMF and Bogoliubov approximation for homogeneous system}
\label{subsec-Comparison}
A large $N$ limit of the IZMF is an interesting aspect to be explored. The numerical
calculation for the inhomogeneous system trapped by a confining potential and 
with large $N$ is a heavy load, thus we study a homogeneous system instead, described by
the Hamiltonian (\ref{eq:Hamiltonian}) without the trapping harmonic potential, 
for a large $N$\,. It is straightforward to develop the IZMF for the homogeneous system.
When the zero mode sector is dropped, we have the well-known formulation of a homogeneous
system in the Bogoliubov approximation.

We assume a system with finite $N$ and $V$ and apply the homogeneous IZMF as above
to it and increase $N$ at $T/T_c=0.01$, keeping the density $n=N/V$ constant.
The ratios of the depletion and specific heat in the homogeneous IZMF to those in the Bogoliubov
approximation versus $N$ are shown in Fig.~\ref{fig:thermoDynamicial}.
It is confirmed that both the ratios converge to unity at a large $N$ limit, which implies 
that the IZMF and the ordinary formulation of the Bogoliubov approximation cannot be 
distinguished from each other in the thermodynamical limit. On the other hand, the ratios 
are rather large for small $N$ and the zero mode effects are clear then.

\section{Summary}\label{sect:Summary}

In this paper, the unperturbed 
formulation for an inhomogeneous condensate system 
at finite temperature beyond the Bogoliubov approximation, called the IZMF, 
is given. While the zero mode operators are being treated properly, 
the infrared divergence, inevitable in the conventional formulation, 
is regularized by including the zero mode interaction into the unperturbed Hamiltonian. 
Then, a unique and stationary density matrix is defined, 
which allows us to calculate observables at thermal equilibrium. 

The whole energy spectrum of the inhomogeneous condensate system consists of 
two parts, the BdG discrete spectrum, also existing in the Bogoliubov approximation,
and the zero mode discrete spectrum 
equivalent to a one-dimensional quantum mechanical bound system.
The level spacings of the two spectra and a typical thermal energy 
$k_\mathrm{B}T$ are key parameters. For realistic experiments in confined
cold atomic gas, the spacing of the zero mode energy varies from being
comparable with that of the BdG mode in stronger interaction to
being much smaller in weaker interaction.  
The zero mode excitation in weaker interaction
has distinctive effects on the depletion and
specific heat at the unperturbed level.
Although we cannot tell much about the condensate fraction
and critical temperature shift\cite{Tc1, Tc2} that were observed at higher temperuture
and for stronger interaction, our results at the unpertubed level 
does not contradict with the experimentally observed negative shift 
of the critical temperature.

Observing the zero mode operators $\Qh$ and $\Ph$ that are canonical variables,
we have noticed that their uncertainty does not reach its minimum value even in very weak
interactions because of the very low excitation energies 
of the zero mode. The spread of the uncertainty may be observed, {\it e.g.} directly 
from the visibility of an interference fringe, or indirectly from an enhancement in the
specific heat, if a condensate formation of very weakly interacting atoms
in thermal equilibrium is achieved despite experimental 
difficulties.

It is an interesting question how the IZMF for the inhomogeneous system and the widely-used
formulation in the Bogoliubov approximation for the homogeneous system 
are related to each other. As discussed in Subsect.~\ref{subsec-Comparison}, 
the quantities in the IZMF approach the calculated values of the Bogoliubov approximation
in a large $N$ limit or in the thermodynamical limit for a homogeneous system.
This verifies the validity of the Bogoliubov approximation for homogeneous
systems, but we stress here that the Bogoliubov approximation is not valid 
for inhomogeneous systems, such as trapped cold atomic gas.

Although this study is restricted to the quantities over 
the unperturbed density matrix, it is possible to perform a systematic calculation of
 higher order perturbation in future. For this, we emphasize that our unperturbed
formulation for a finite-size system satifsies all the requirements of quantum field theory,
including the canonical commutation relations, and is free from the infrared divergence.

\section*{Acknowledgements}
This work is supported in part by JSPS KAKENHI Grant No.~25400410 and No.~16K05488.

% % % % % % % % % % % % % % % % % % % % % % % % % % %
\section*{References}
\bibliography{reference}

\end{document}